\documentclass [a4paper,fleqn, 12pt]{article}
\usepackage{graphicx}
\usepackage[small]{subfigure,epsfig}

\usepackage {amsmath} \usepackage{amssymb} \usepackage{cite}

\begin{document}

\title{Analytical solutions of the Rayleigh equation for empty and gas--filled bubble}

\author{Nikolay A. Kudryashov and Dmitry I. Sinelshchikov}
\date{Department of Applied Mathematics, National Research Nuclear University MEPhI (Moscow Engineering Physics Institute), 31 Kashirskoe Shosse, 115409 Moscow, Russian Federation}

\maketitle

\begin{abstract}
The Rayleigh equation for bubble dynamics is widely used. However, analytical solutions of this equation have not been obtained previously. Here we find closed--form general solutions of the Rayleigh equation both for an empty and gas--filled spherical bubble. We present an approach allowing us to construct exact solutions of the Rayleigh equation. We show that our solutions are useful for testing numerical algorithms.
\end{abstract}

\section{Introduction}
\indent
Lord Rayleigh derived the equation for the dynamics of a gas--filled cavity in an incompressible liquid neglecting liquid viscosity, surface tension and thermal effects \cite{Rayleigh1917}. Then this equation was generalized by M.S. Plesset \cite{Plesset1977} to include liquid viscosity and surface tension. Now the Rayleigh equation and its generalizations are widely used for studying various phenomena where gas bubbles dynamics is important \cite{Brennen2013,Bogoyavlenskiy2000,Lauterborn2010,Kudryashov2010a,Doinikov2013,Kudryashov2013}.

Usually dynamics of a gas bubble governed by the Rayleigh equation is studied numerically (see, e.g. \cite{Bogoyavlenskiy2000,Lauterborn2010,Doinikov2013}). Recently, accurate asymptotic solutions of the Rayleigh equation were constructed \cite{Obreschkow2012,Fernandez2013}. In previous works (see, e.g. \cite{Alehossein2007,Obreschkow2012}) it was noted that there are no closed--form solutions of the Rayleigh equation. However, analytical solutions can be useful both for the investigation of bubbles dynamics and for testing numerical algorithms for solving the Cauchy problem for the Rayleigh equation.

Here we present an approach allowing us to construct analytical solutions of the Rayleigh equation both for the case of an empty and gas--filled bubble. Using this approach we find several analytical solutions of the Rayleigh equation which are expressed via hypergeometric and elliptic functions. We demonstrate that obtained analytical solutions can be used for testing of numerical algorithms.

\section{Main equations and approach}
Let us consider both the Rayleigh equation for an empty bubble
\begin{equation}
\rho\left(R R_{TT}+\frac{3}{2}R_{T}^{2}\right)=-p,
\label{eq:Rayleigh_empy_cavity}
\end{equation}
and the Rayleigh equation for a gas--filled bubble
\begin{equation}
\rho\left(R R_{TT}+\frac{3}{2}R_{T}^{2}\right)=P_{0}\left(\frac{R_{0}}{R}\right)^{3\kappa}-p,
\label{eq:Rayleigh_gas_filled_cavity}
\end{equation}
where $R$ is the radius of the bubble, $T$ is the time, $p$ is the far--field pressure, $\rho$ is the liquid density, $\kappa$ is the polytropic exponent, $P_{0}$ is the ambient pressure of the gas in the bubble,  $R_{0}$ is the ambient radius of bubble. Throughout this work we use subscripts to denote derivatives.  We assume that the far--field pressure is constant.   In \eqref{eq:Rayleigh_gas_filled_cavity} we also suppose that the gas in the bubble is an ideal and obeys the polytropic law.

Using the same non--dimensional variables as in \cite{Obreschkow2012} $R=R_{0}u$, $T=T_{c}t$, $T_{c}^{2}=\xi^{2}R_{0}^{2}\rho/p$ from \eqref{eq:Rayleigh_empy_cavity} we get
\begin{equation}
u u_{tt}+\frac{3}{2}u_{t}^{2}+\xi^{2}=0,
 \label{eq:Reley_eq_ec_non_dim}
\end{equation}
where $\xi=\int_{0}^{1}(r^{-3}-1)^{-1/2}dr\approx 0.914681$ is a universal constant called the Rayleigh factor and $T_{c}$ is the collapse time.

We use the following non--dimensional variables $T=\omega_{0}^{-1}\,t$, $R=R_{0}u$  in \eqref{eq:Rayleigh_gas_filled_cavity} to obtain
\begin{equation}
u u_{tt}+\frac{3}{2} u_{t}^{2}=\frac{1}{3\kappa}\left[u^{-3\kappa}-\beta\right],
\label{eq:Reley_eq_nec_non_dim}
\end{equation}
where $\beta=p/P_{0}$ and $\omega_{0}^{2}=3\kappa P_{0}/(\rho R_{0}^{2})$ is the natural frequency of bubble oscillations.

Multiplying \eqref{eq:Reley_eq_ec_non_dim} and \eqref{eq:Reley_eq_nec_non_dim} by $2u^{2}u_{t}$ and integrating the results with respect to $t$ we have
\begin{equation}
u_{t}^{2}=C_{1}u^{-3}-\frac{2}{3}\xi^{2},
 \label{eq:Reley_eq_ec_non_dim_fi}
\end{equation}
\begin{equation}
u_{t}^{2}=\frac{2}{9\kappa(1-\kappa)}u^{-3\kappa}-\frac{2\beta}{9\kappa}+C_{1}u^{-3},
\label{eq:Reley_eq_nec_non_dim_fi}
\end{equation}
where $C_{1}$ is an integration constant. Let us note that we do not consider the isothermal case ($\kappa=1$) in this work. Let us also remark that for finding first integrals \eqref{eq:Reley_eq_ec_non_dim_fi}, \eqref{eq:Reley_eq_nec_non_dim_fi} we can use an approach from Kamke's book (see \cite{Kamke}, case 6.165). 

Note that the physically possible range of the polytropic exponent for practical bubble dynamics is $1<\kappa\leq5/3$ (recall that we do not consider the case of $\kappa=1$). Consequently, the first two terms in the right--hand side of \eqref{eq:Reley_eq_nec_non_dim_fi} are negative and it immediately follows that $C_{1}>0$ for solutions of \eqref{eq:Reley_eq_nec_non_dim_fi} to be real. We can also consider \eqref{eq:Reley_eq_nec_non_dim_fi} as the energy conservation law. Therefore, $C_{1}$ represents the non--dimensional total energy and thus in our case it is greater than zero.

Now our main goal is to find general solutions of \eqref{eq:Reley_eq_ec_non_dim_fi}, \eqref{eq:Reley_eq_nec_non_dim_fi}. It seems impossible to construct solutions of \eqref{eq:Reley_eq_ec_non_dim_fi}, \eqref{eq:Reley_eq_nec_non_dim_fi} by their direct integration. However, it is possible to find some transformations which allow us to convert each of these equations into an equation with a known general solution. We consider transformations of \eqref{eq:Reley_eq_ec_non_dim_fi}, \eqref{eq:Reley_eq_nec_non_dim_fi} into a first order nonlinear ordinary differential equation of second degree with a polynomial in the right--hand side. Among equations of this type, except the linearized equation with a quadratic polynomial, only equations with a cubic and quatric polynomials in the right--hand side have single valued solutions \cite{Hille}. They are equations for the Weierstrass and Jacobi elliptic functions \cite{Hille,Whittaker}.  We will transform each of \eqref{eq:Reley_eq_ec_non_dim_fi}, \eqref{eq:Reley_eq_nec_non_dim_fi} into one of these equations. To this end, we use the Sundman transformation combined with a power--type transformation:
\begin{equation}
dt=u^{\delta}\,d\tau, \quad u=v^{\epsilon}.
\label{eq:main_transformations}
\end{equation}
Here $\tau$ and $v$ are new independent and dependent variables correspondingly, $\delta$ and $\epsilon\neq0$ are real numbers

Substituting \eqref{eq:main_transformations} into each of \eqref{eq:Reley_eq_ec_non_dim_fi}, \eqref{eq:Reley_eq_nec_non_dim_fi} we obtain two first order second degree nonlinear ordinary differential equations with a rational function in the right--hand side.
Consider the case of equation \eqref{eq:Reley_eq_ec_non_dim_fi}. Under requirement that the resulting equation is one of the equations for elliptic functions we obtain a system of two algebraic equations for parameters $\delta$ and $\epsilon$.
In the same way we obtain a system of three algebraic equations for parameters $\delta$, $\epsilon$ and $\kappa$ in the case of \eqref{eq:Reley_eq_nec_non_dim_fi}. Details of application of \eqref{eq:main_transformations} to \eqref{eq:Reley_eq_ec_non_dim_fi} and \eqref{eq:Reley_eq_nec_non_dim_fi} will be presented in \ref{sec3} and \ref{sec4} correspondingly. It is worth noting that differential equations for the Weierstrass and Jacobi elliptic functions are connected to each other. Consequently, one can choose any of these equations as a resulting equation.

Let us note that the Sundman transformation was proposed in work \cite{Sundman1913} for studying the 3--body problem and later this transformation was used in studies of nonlinear ordinary differential equation and in celestial mechanics (see, eg. \cite{Duarte1994,Meleshko2010,Nucci2010,Meleshko2011}).

\section{\label{sec3} The case of empty bubble}

Let us construct the general solution of the Rayleigh equation for the motion of the empty spherical bubble. Applying \eqref{eq:main_transformations} with $\delta=4$ and $\epsilon=1/3$ to \eqref{eq:Reley_eq_ec_non_dim_fi} we get
\begin{equation}
v_{\tau}^{2}=9C_{1}v^{3}-6\xi^{2}v^{4}.
 \label{eq:ec_5}
\end{equation}
The general solution of \eqref{eq:ec_5} has the form
\begin{equation}
v=\frac{12C_{1}}{27C_{1}^{2}(\tau-\tau_{0})^{2}+8\xi^{2}}
 \label{eq:ec_5_a}
\end{equation}
Using \eqref{eq:ec_5_a} and \eqref{eq:main_transformations} we obtain the general solution of the Rayleigh equation for the empty spherical bubble:
\begin{equation}
u=\left[\frac{12C_{1}}{27C_{1}^{2}(\tau-\tau_{0})^{2}+8\xi^{2}}\right]^{1/3}, \quad
t=\int\limits_{0}^{\tau}u^{4}(\zeta)d\,\zeta.
 \label{eq:ec_9}
\end{equation}
We denote by $\zeta$ and $\tau_{0}$ a dummy integration variable and an integration constant correspondingly throughout this work. The integral in \eqref{eq:ec_9} is expressed via the hypergeometric function. Taking into account this fact and solving the first equality from \eqref{eq:ec_9} for $\tau$ we find the dependence of $t$ on $u$:
\begin{equation}
\begin{gathered}
\pm t=\left(\frac{3C_{1}}{2\xi^{2}}\right)^{\frac{4}{3}}
\Bigg[\tau_{0}F\left\{\frac{1}{2},\frac{4}{3};\frac{3}{2};-\frac{27C_{1}^{2}}{8\xi^{2}}\tau_{0}^{2}\right\}+\\
+\frac{2}{3\sqrt{C_{1}}}\left(\frac{1}{u^{3}}-\frac{2\xi^{2}}{3C_{1}}\right)^{1/2}
F\left\{\frac{1}{2},\frac{4}{3};\frac{3}{2};\frac{C_{1}3}{2\xi^{2}}\left[\frac{2\xi^{2}}{3C_{1}}-\frac{1}{u^{3}}\right]\right\}\Bigg],
 \label{eq:emty_cavity_5_a}
\end{gathered}
\end{equation}
where $F$ is the hypergeometric function and the sign $\pm$ corresponds to the invariance of \eqref{eq:Reley_eq_ec_non_dim} under the transformation $t\rightarrow -t$. Consequently, we find the general closed--form solution of the Rayleigh equation for the empty spherical bubble.

\begin{figure}[!t]
\center
\includegraphics[width=6.5cm,height=5cm]{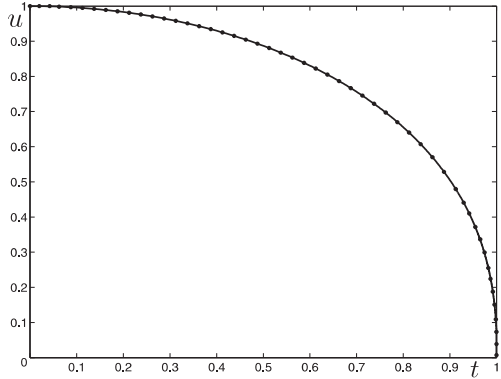}
\caption{Exact solution \eqref{eq:ec_17} of \eqref{eq:Reley_eq_ec_non_dim} (shown by solid curve) and numerical solution of the Cauchy problem \eqref{eq:Reley_eq_ec_non_dim},\eqref{eq:ec_11} (shown by dotted curve).}
\label{f7}
\end{figure}

Let us consider the Cauchy problem for \eqref{eq:Reley_eq_ec_non_dim} with initial conditions corresponding to the collapse motion of the empty spherical bubble \cite{Obreschkow2012}:
\begin{equation}
u(0)=1,\quad u_{t}(0)=0.
 \label{eq:ec_11}
\end{equation}
Taking into consideration \eqref{eq:ec_11} we find that $C_{1}=2\xi^{2}/{3}$ and $\tau_{0}=0$. Using these values of $C_{1}$ and $\tau_{0}$ in \eqref{eq:ec_9} we obtain the exact solution of \eqref{eq:Reley_eq_ec_non_dim} corresponding to \eqref{eq:ec_11} in the parametric form:
\begin{equation}
u=\left[\frac{2}{3\xi^{2}\tau^{2}+2}\right]^{1/3}, \quad
t=\int\limits_{0}^{\tau}u^{4}(\zeta)d\,\zeta.
 \label{eq:ec_17}
\end{equation}
Using \eqref{eq:emty_cavity_5_a} we also can find the closed--form solution of problem \eqref{eq:Reley_eq_ec_non_dim},\eqref{eq:ec_11}:
\begin{equation}
\pm t=\sqrt{\frac{2}{3}}\frac{1}{\xi}\left[\frac{1}{u^{3}}-1\right]^{1/2} F\left\{\frac{1}{2},\frac{4}{3};\frac{3}{2};1-\frac{1}{u^{3}}\right\}.
\label{eq:ec_15}
\end{equation}
The asymptotic expansion for \eqref{eq:ec_15} has the form
\begin{equation}
t=1-\frac{\sqrt{6}}{5\xi}u^{5/2}-\frac{\sqrt{6}}{22\xi}u^{11/2}-\frac{3\sqrt{6}}{136\xi}u^{17/2}+O\left(u^{23/2}\right).
\label{eq:ec_15a}
\end{equation}
The same rapidly convergent asymptotic expansion was obtained in \cite{Fernandez2013}.

Solution \eqref{eq:ec_17} (or \eqref{eq:ec_15}) is shown in Figure \ref{f7}. We can see that this solution describes the collapsing spherical bubble. Solution \eqref{eq:ec_17} qualitatively agrees with asymptotical and experimental results of work \cite{Obreschkow2012}. We also present numerical solution of Cauchy problem \eqref{eq:Reley_eq_ec_non_dim},\eqref{eq:ec_11} in Figure \ref{f7}.  We can see a good agreement between analytical and numerical solutions. Thus, one can use solution \eqref{eq:ec_17} for testing programs for numerical solving of the Cauchy problem for \eqref{eq:Reley_eq_ec_non_dim}. Note that throughout this work we use the Cash-Karp fourth-fifth order Runge--Kutta method \cite{Obreschkow2012}.

\begin{figure}[!t]
\center
\includegraphics[width=6.5cm,height=5cm]{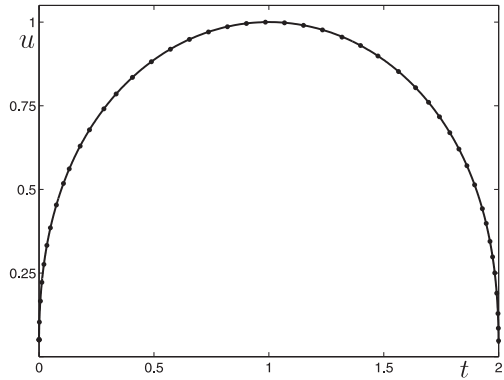}
\caption{Exact solution \eqref{eq:ec_17} of \eqref{eq:Reley_eq_ec_non_dim} corresponding to initial conditions \eqref{eq:ec_19} (shown by solid line) and numerical solution of Cauchy problem \eqref{eq:Reley_eq_ec_non_dim}, \eqref{eq:ec_19} .}
\label{f8}
\end{figure}

Let us suppose that $C_{1}$ has the same value as in the previous case and $\tau_{0}=\sqrt{47994}/(3\xi)$. In this case we have the following initial conditions
\begin{equation}
u(0)=0.05, \quad u_{t}(0)=\tau_{0}\xi^{2}=\frac{\sqrt{47994}\xi}{3}.
 \label{eq:ec_19}
\end{equation}
Solution \eqref{eq:ec_9} corresponding to \eqref{eq:ec_19} describes growth and collapse of the empty spherical bubble and is shown in Figure \ref{f8}. We can see that this solution qualitatively agrees with experimental results of work \cite{Obreschkow2013}. We also present numerical solution of Cauchy problem \eqref{eq:Reley_eq_ec_non_dim}, \eqref{eq:ec_19} in Figure \ref{f8}. We can see a good agreement between analytical and numerical results.

In this section we have found the general closed--form solution of the Rayleigh equation for the empty spherical bubble. We have demonstrated that our solutions qualitatively agrees with experimental data.

\section{\label{sec4}The case of gas--filled bubble}

\begin{figure}[!tp]
\center
\includegraphics[width=6.5cm]{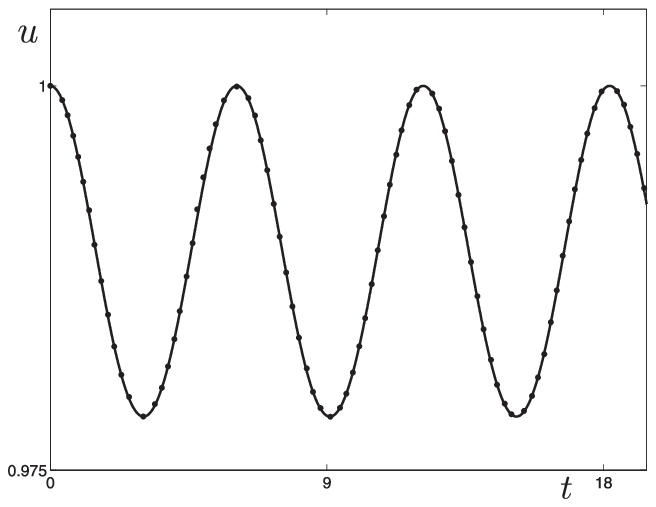}
\caption{Periodic exact solution \eqref{eq:R_29} of equation \eqref{eq:Reley_eq_nec_non_dim} at $\kappa=3/2$ corresponding to initial conditions \eqref{eq:R_31} (shown by solid curve) and numerical solution of the Cauchy problem  \eqref{eq:Reley_eq_nec_non_dim}, \eqref{eq:R_31}  (shown by dotted curve). }
\label{f1}
\end{figure}

In this section we consider the case of the gas-filled bubble. Let us recall that applying transformations \eqref{eq:main_transformations} to \eqref{eq:Reley_eq_nec_non_dim_fi} and requiring that the resulting equation will be one of the equations for elliptic functions we obtain a system of three algebraic equations for parameters $\delta$, $\epsilon$ and $\kappa$. Supposing that $\kappa$ is a positive number and solving this system of equations we find that $\kappa$ may take the following values: $1/3$, $2/3$, $4/3$, $1/4$, $3/4$, $1/2$, $3/2$, $2$, $3$, $4$. Let us remark that physically possible values of $\kappa$ are $\kappa=3/2$ and $\kappa=4/3$. The value of $\kappa=3/2$ corresponds to behaviour  of helium, neon, argon, krypton or radon between isothermal and adiabatic. Note that this is a rare case. The value of $\kappa=4/3$ corresponds to behaviour of diatomic gases between isothermal and adiabatic which is the most realistic case. Below we consider these two values of $\kappa$.

Let us suppose that $\kappa=3/2$. First, applying \eqref{eq:main_transformations} with $\delta=5/2$ and $\epsilon=-2/3$ we get the equation
\begin{equation}
v_{\tau}^{2}=-\frac{2}{3}v^{3}+\frac{9C_{1}}{4}v^{2}-\frac{\beta}{3}.
\label{eq:R_25}
\end{equation}

Then, we find the general solution of \eqref{eq:R_25}
\begin{equation}
\begin{gathered}
v=\frac{9C_{1}}{8}-\wp\left\{\frac{1}{\sqrt{6}}(\tau-\tau_{0}),g_{2},g_{3}\right\}, \\
g_{2}=3\left(\frac{9C_{1}}{4}\right)^{2}, \quad g_{3}=2\beta-\left(\frac{9C_{1}}{4}\right)^{3},
\label{eq:R_27}
\end{gathered}
\end{equation}
where $\wp$ is the Weierstrass elliptic function and $g_{2}$, $g_{3}$ are the invariants of the Weierstrass elliptic function. And finally, using \eqref{eq:main_transformations} we obtain the general solution of \eqref{eq:Reley_eq_nec_non_dim} at $\kappa=3/2$ in the parametric form
\begin{equation}
u=\left[\frac{9C_{1}}{8}-\wp\left\{\frac{1}{\sqrt{6}}(\tau-\tau_{0}),g_{2},g_{3}\right\}\right]^{-2/3},\quad
t=\int\limits_{0}^{\tau} u^{5/2}(\zeta) d\zeta\,. \hfill
\label{eq:R_29}
\end{equation}
Solutions \eqref{eq:R_27} and \eqref{eq:R_29} have poles on the real axis in the case of $\tau_{0}=0$. However, assuming that the equation $4w^{3}-g_{2}w-g_{3}$ has three distinct real roots we can remove poles from the real line. Supposing that $\tau_{0}=-\sqrt{6}\omega_{3}$, where $\omega_{3}$ is the imaginary half--period of the Weierstrass elliptic function, and using special case of additional theorem (e.g., see \cite{Whittaker}) we obtain a solution bounded on the real line. In what follows, it will be assumed that the transformation $\tau_{0}\rightarrow\tau_{0}-\sqrt{6}\omega_{3}$ was made in solution \eqref{eq:R_29}.

\begin{figure}[!tp]
\center
\includegraphics[width=6.5cm]{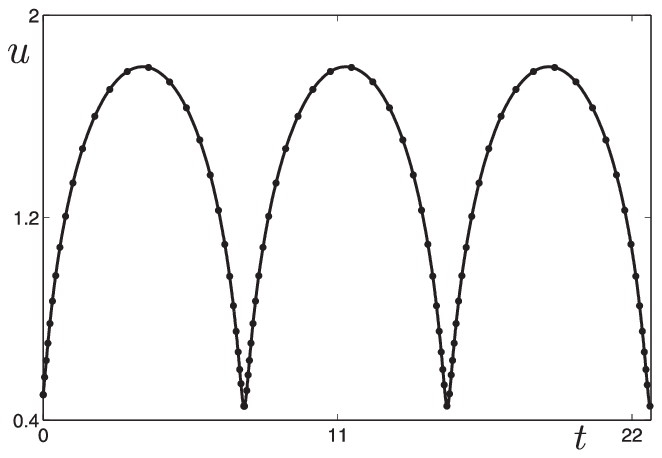}
\caption{Periodic exact solution \eqref{eq:R_29} of equation \eqref{eq:Reley_eq_nec_non_dim} at $\kappa=3/2$ corresponding to initial conditions \eqref{eq:R_33} (shown by solid curve) and numerical solution of the Cauchy problem  \eqref{eq:Reley_eq_nec_non_dim}, \eqref{eq:R_33}  (shown by dotted curve).}
\label{f2}
\end{figure}

It is worth noting that as far as $\beta>0$ and $C_{1}>0$ solution \eqref{eq:R_29} always has a real period. Consequently, the motion of the gas--filled bubble described by \eqref{eq:Reley_eq_nec_non_dim} at $\kappa=3/2$ can be periodic only.

Let us consider the Cauchy problem for \eqref{eq:Reley_eq_nec_non_dim} with the following initial conditions
\begin{equation}
u(0)=1, \quad u_{t}(0)=0,
\label{eq:R_31}
\end{equation}
and the parameter $\beta=1.05$. This value of $\beta$ corresponds to a 5\% steep change in the far--field pressure from the equilibrium pressure $P_{0}$ at $t=0$. Using \eqref{eq:R_31} and \eqref{eq:Reley_eq_nec_non_dim_fi} we find that $C_{1}=62/135$. As soon as we know parameters $\beta$ and $C_{1}$ we can calculate the invariants $g_{2}$, $g_{3}$. Taking into account values of $\beta$ and $C_{1}$ and the second condition from \eqref{eq:R_31} we find that $\tau_{0}$ is a solution of the equation $\wp_{t}\{1/\sqrt{6}\tau_{0},g_{2},g_{3}\}=0$, which can be easily obtained using a symbolic software package such as Maple or Mathematica. The plot of solution \eqref{eq:R_29} corresponding to \eqref{eq:R_31} is presented in Figure \ref{f1}. We can see that this solution represents periodic oscillations of the bubble under influence of the constant pressure. Numerical solution of Cauchy problem \eqref{eq:Reley_eq_nec_non_dim}, \eqref{eq:R_31} is also shown in Figure \ref{f1}. We can see a good agreement between analytical and numerical results. Thus, this exact solution can be used for testing programs for numerical solving of the Cauchy problem for the Rayleigh equation. Using advantages of an analytical approach we can find the exact value of the period of the solution presented in Figure \ref{f1}. It has the form $\sqrt{6}\int_{e_{1}}^{\infty}dw/\sqrt{(w-e_{1})(w-e_{2})(w-e_{3})} \approx 6.2321$, where $e_{1},e_{2},e_{3}$ are roots of the equation $4w^{3}-g_{2}w-g_{3}$.

Now we consider the following initial conditions
\begin{equation}
u(0)=0.5, \quad u_{t}(0)=1,
\label{eq:R_33}
\end{equation}
which correspond to the grow motion of the bubble at $t=0$.  We also suppose that $\beta=1$. Taking into account \eqref{eq:Reley_eq_nec_non_dim_fi} and using \eqref{eq:R_33} we find that $C_{1}=(16\sqrt{2}+31/8)/27$. One can find the value of $\tau_{0}$ in the same way as in the previous case. We demonstrate solution \eqref{eq:R_29} corresponding to \eqref{eq:R_33} in Figure \ref{f2}. We see that this solution describes the rapid growth and compression motion of the bubble. We also demonstrate in Figure \ref{f2} numerical solution of Cauchy problem \eqref{eq:Reley_eq_nec_non_dim}, \eqref{eq:R_33}. We can see a good agrement between analytical and numerical results.

Now let us consider the case of $\kappa=\frac{4}{3}$. Using \eqref{eq:main_transformations} with $\delta=2$ and $\epsilon=-1$ we obtain the equation
\begin{equation}
v_{\tau}^{2}=-\frac{1}{2}v^{4}+C_{1}v^{3}-\frac{\beta}{6}.
\label{eq:R_35}
\end{equation}
The general solution of \eqref{eq:R_35} can be expressed via one of the Jacobi elliptic functions. This solution can be found using a standard approach (see, e.g. \cite{Whittaker}), however, it has a rather cumbersome form. On the other hand, we can find expression for the general solution of \eqref{eq:R_35} in terms of the Weierstrass elliptic function which has a simple form. Indeed, let the parameter $\mu$ be a real solution of the following equation
\begin{equation}
\beta \mu^{4}-48(C_{1}\mu-1)=0,
\label{eq:R_37}
\end{equation}
then the general solution of \eqref{eq:R_35} has the form
\begin{equation}
\begin{gathered}
  v= \frac{ 4\mu^{2}\,\wp \{ \tau+\tau_{0},g_{2},g_{3} \} + 4C_{1}\mu-4
 }  {2\mu^{3}\wp \{ \tau+ \tau_{0},g_{2},g_{3} \} -\mu(
C_{1}\mu-2)} ,\\
g_{2}=\frac {4(C_{1}\mu-1)}{\mu^{4}}=\frac{\beta}{12}, \quad g_{3}=\frac {C_{1}^{2} ( C_{1}
\mu-1 ) }{2\mu^{4}}=\frac{\beta C_{1}^{2}}{96}, \hfill
\label{eq:R_39}
\end{gathered}
\end{equation}
where we again denote by $\wp$ the Weierstrass elliptic function and by $g_{2}$, $g_{3}$ its invariants.

Using \eqref{eq:main_transformations} and \eqref{eq:R_39} we find the general solution of \eqref{eq:Reley_eq_nec_non_dim} at $\kappa=4/3$ in the parametric form
\begin{equation}
 u= \frac{2\mu^{3}\wp \{ \tau+ \tau_{0},g_{2},g_{3} \} -\mu(
C_{1}\mu-2)}{ 4\mu^{2}\,\wp \{ \tau+\tau_{0},g_{2},g_{3} \} + 4C_{1}\mu-4
 }    ,\quad
t=\int\limits_{0}^{\tau} u^{2}(\zeta) d\zeta\,. \hfill
\label{eq:R_41}
\end{equation}
It is worth noting that solution \eqref{eq:R_41} has no poles on the real axis at $\beta>0$ and $C_{1}>0$.

Let us remark that as far as $\beta>0$ and $C_{1}>0$ solutions of \eqref{eq:R_35} have a real period. Consequently, the motion of the gas--filled bubble described by \eqref{eq:Reley_eq_nec_non_dim} at $\kappa=4/3$ is periodic. Note that as we have shown above the same is true for the case of $\kappa=3/2$.

\begin{figure}[!tp]
\center
\includegraphics[width=6.5cm]{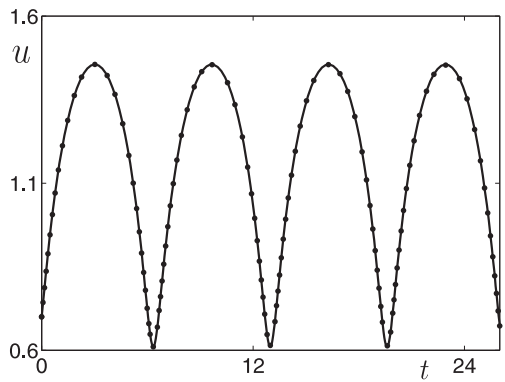}
\caption{Periodic exact solution \eqref{eq:R_41} of equation \eqref{eq:Reley_eq_nec_non_dim} at $\kappa=4/3$ corresponding to initial conditions \eqref{eq:R_43} (shown by solid curve) and numerical solution of the Cauchy problem  \eqref{eq:Reley_eq_nec_non_dim}, \eqref{eq:R_43}  (shown by dotted curve).}
\label{f5}
\end{figure}

Now let us consider a solution of \eqref{eq:Reley_eq_nec_non_dim} at $\kappa=4/3$ corresponding to some particular initial conditions. We assume that $\beta=1$ and the following relations hold
\begin{equation}
u(0)=0.7, \quad u_{t}(0)=0.5.
\label{eq:R_43}
\end{equation}
These initial conditions correspond to the growth motion of the gas--filled bubble. Taking into account \eqref{eq:R_43} and using \eqref{eq:Reley_eq_nec_non_dim_fi} we find that $C_{1}=14401/16800$. Since we know parameters $\beta$ and $C_{1}$ we can find values of the parameter $\mu$ and invariants $g_{2}$, $g_{3}$. Then we can obtain a value of $\tau_{0}$ in the same way as in the case of $\kappa=3/2$. Solution \eqref{eq:R_41} corresponding to \eqref{eq:R_43} is shown in Figure \ref{f5}. We can see that this solution describes smooth growth and compression motion of the gas--filled bubble. Numerical solution of Cauchy problem \eqref{eq:Reley_eq_nec_non_dim}, \eqref{eq:R_43} is also demonstrated in Figure \ref{f5}. We can see a good agreement between analytical and numerical solutions.

Let us finally remark that we have also considered solutions of Cauchy problems \eqref{eq:Reley_eq_nec_non_dim}, \eqref{eq:R_31} and \eqref{eq:Reley_eq_nec_non_dim}, \eqref{eq:R_33} at $\kappa=4/3$. We have found that the bubble motion in this case is similar to the bubble motion in the case of $\kappa=3/2$.

In this section we have given the general solution of the Rayleigh equation for the gas--filled bubble in the case of the polytropic exponent equal to $3/2$. We have also found the general solution of the Rayleigh equation for the case of $\kappa=4/3$.

\section{Conclusion}
We have considered the Rayleigh equation for bubble dynamics. We have proposed an approach allowing us to construct analytical solutions of the Rayleigh equation both in the case of the empty and gas--filled bubble. We have obtained the general solution of the Rayleigh equation in the case of the empty spherical bubble. We have presented two special exact solutions of the Rayleigh equation that correspond to the collapse and growth and collapse motion of the empty spherical bubble. We have shown that the general solution of the Rayleigh equation in the case of the gas--filled bubble can be constructed for certain values of the polytropic exponent. Among these values of the polytropic exponent only two are physically relevant. They are $\kappa=3/2$ and $\kappa=4/3$. We have considered the case of $\kappa=3/2$ and presented the general solution of the Rayleigh equation that is expressed via the Weierstrass elliptic function. We have also found the general solution of the Rayleigh equation for the case of $\kappa=4/3$ which is again expressed via the Weierstrass elliptic function. At the best of our knowledge our solutions are new. We have demonstrated that our solutions can be used for testing of numerical algorithms for solving the Cauchy problem for the Rayleigh equation. We have compared some of our solutions with experimental data \cite{Obreschkow2012,Obreschkow2013} and found qualitative agreement between the former and the latter. We have also shown that the asymptotic approximation for the collapse motion of the empty spherical bubble can be obtained from the corresponding analytical solution.

Authors are grateful to anonymous referees for their valuable comments and suggestions.

This research was supported in part by RFBR grant 14-01-00493-a, by grant for Scientific Schools 2296.2014.1 and by grant for the state support of young Russian scientists 3694.2014.1.


\end{document}